# Laying the Groundwork for a Worker-Centric Peer Economy

**Ali Alkhatib**
Stanford University
Stanford, USA
ali.alkhatib@cs.stanford.edu

**Justin Cranshaw**
Microsoft Research
Redmond, WA, USA
justincr@microsoft.com

**Andrés Monroy-Hernández**
Microsoft Research
Redmond, WA, USA
andresmh@microsoft.com

## ABSTRACT
The "gig economy" has transformed the ways in which people work, but in many ways these markets stifle the growth of workers and the autonomy and protections that workers have grown to expect. We explored the viability of a "worker-centric peer economy"—a system wherein workers benefit as well as consumers— and conducted ethnographic field work across fields ranging from domestic labor to home health care.

We discovered seven facets that system designers ought to consider when designing a labor market for "gig workers," consisting principally of the following: constructive feedback, assigning work fairly, managing customer expectations, protecting vulnerable workers, reconciling worker identities, assessing worker qualifications, & communicating worker quality. We discuss these considerations and provide guidance toward the design of a mutually beneficial market for gig workers.

## ACM Classification Keywords
H.5.m. Information Interfaces and Presentation (e.g. HCI): Miscellaneous; See <http://acm.org/about/class/1998/> for the full list of ACM classifiers. This section is required.

## INTRODUCTION
In the past several years, a new type of work has emerged where customers can hire someone to complete an individual task, or "gig." This task might involve delivering a package, or driving someone downtown. Such a model principally requires the sharing of capital-intensive goods, like access to a car or home, which led to the popularization of its familiar name, the "sharing economy." People could "share" their homes (Airbnb, Couchsurfing [6, 3]), cars (Uber, Lyft, and others [5, 2]), and increasingly one's own time (TaskRabbit, Zaarly, and many others [4, 1]).

Workers' demographics have changed dramatically in as little as half a decade, largely discreetly. Where workers in the sharing economy were once car and home-owners who had free time to spare, many now think of these companies and the markets they expose as their primary source of income. In the last several years, workers for Uber, Homejoy, and other market operators have initiated —and in some cases won [19, 15]— suits describing mistreatment and misclassification of these workers as "independent contractors" rather than "employees" (protected and regulated by labor laws in the United States) [13].

In the sharing economy's nascent years, companies enticed workers to join for the potential to do work "on the side": offering rides to others when they had free time, or renting out their apartment when they were out of town for a weekend. For various reasons, the culture has since changed, and the notion of doing work in one's free time has largely disappeared. Instead, drivers report primarily working as drivers, and that their primary sources of income consist of the aggregated sum of passengers they pick up through ride-sharing markets. Even in the hotelier industry, providers have purchased apartments or re-purposed their own homes primarily to serve guest occupants, rather than to rent out incidentally when they have a spare room or are out of town.

In these ways, workers are neither "peers," nor are they "sharing" resources that would otherwise go underutilized. But they are not conventional workers, either. They acquire capital —sometimes co-signing on leases with the companies that run these markets— under their own names, run their businesses relatively independently, and make the majority or even totality of their income based on each individual job, or "gig," cumulatively summed up. With their careers described as a series of individual jobs, each self-contained and relatively independent of the others, people have renamed it the "gig economy," more fairly referencing the differentiating nature of this work.

The widespread nature of these changes suggests that this is part of a larger trend in what may become the future of work; far from the hopeful but cautious predictions offered of information workers and "crowd work" [21]. Workers increasingly find themselves objectified, marginalized, and frustrated by oppressive systems.

We considered, then, how one might design a worker-centric peer market; how can system designers create technologically enabled markets as successful as existing markets like Uber, Lyft, and others, while also:

- giving workers a sense of locus;
- benefiting workers as well as consumers;
- facilitating worker organization and communication;

- enabling collective decision & action among workers.

To answer these questions, we engaged in extensive fieldwork alongside workers and labor organizations in a number of industries. Building in part from backgrounds in the social sciences and as trained computer scientists, we learned about workers and the industries in which they work from the workers themselves. We report on the processes of making contact with various formally organized worker advocacy groups, illustrate some of the ways that we can learn from informants most effectively given our own skills, and finally describe some of the findings we made as a result of our own use of these methods.

Informed by the input of dozens of workers from numerous industries ranging from highly regulated to informal, we identify a number of aspects of on-demand work which system designers should consider in the creation of a worker-centric labor market. We offer guidance on these design considerations, and in some cases illustrate the suggested approaches we generated in tandem with these partner organizations and workers.

Specifically, we offer contributions to the following questions

- What components of existing markets are inextricable from the features which make these markets successful?
- What can be disentangled and abstracted away?
- How would these groups be operated?

## BACKGROUND & THEORY

The existing knowledge from which we draw in this research can primarily be traced to two pools of research: The first comes from the extensive body of research surrounding collective action, and the work that has gone into describing, categorizing, predicting, and even designing to foster campaigns of collective action; the second major source of our knowledge comes from the study of on-demand or gig markets. By synthesizing the discoveries from these sources of research, we identify *a priori* guidance at this intersection of two subjects.

### The Long History of Collective Action

Grassroots, community-led organizations are not new: a substantial body of literature illustrates myriad approaches to guiding communities and assisting in collective action. Specifically, when we consider the role of insight into collective action, we refer to action that Hardin describes as "directed at an ongoing problem" [16]. The implication here, he argues, is that the guidance on this form of collective action is dramatically more nuanced than "one-shot" collective action, demanding an "anthropological investigation of minute interrelationships." We might call this "ongoing" collective action. Economist Mancur Olson proposes, in part, that collective action depends on some large, generally inactive group in order to succeed [27].

After decades of observations, Hardin posits that collective action is too commonplace for Olson's thesis to hold; He suggests that the requirements for collective action which Olson theorizes may have changed—specifically, lowering the threshold—as a result of myriad factors outside of the scope of this research, except to point out that recent work in online collective action prompts further scrutiny of Olson's thesis and the critiques later researchers have levied. We suggest an alternative consideration: that much of the research in collective action in the space of human-computer interaction (HCI) in fact corroborates the latent community requirement Olson recommends. Myriad collective action endeavors seem to succeed in part because they precipitate a collective of latent, willing participants in some form of community action [9, 30, 29].

### On-Demand Markets and Their Workers

A robust and growing body of research exploring collective action and movements enabled by the Internet adds to a body of knowledge previously uninformed by the tools the Internet affords. Where collective action research coordinated and executed offline describes challenges symptomatic of social structures, researchers of online communities can and do offer design guidance for the structure of online communities [18]. Substantial contributions deeply investigating online communities, such as Wikipedia, have lent system designers guidance in designing communities geared toward some ongoing collective action online [26, 28, 31]. In this last case, studies of Wikipedia and its users—known colloquially as "Wikipedians"—is especially instructive, as it addresses the distinction Olson makes between "one-shot" collective action and what we will call "ongoing" collective action.

Researchers have described the potential of crowd work, and perhaps more importantly outlined various problems with said markets [21]. More recently, researchers explored the impact of the sharing economy among socioeconomically disadvantaged communities, and studied norms and behaviors among the workers of individual markets like Uber [12, 24]. The directives offered by Kittur et al. provide substantial guidance in the conceptual design of a "crowd market," but the direct application of this research to a communally led organization —a worker-run labor market—merits further consideration.

The focus of technologically enabled peer economies is well-explored; researchers have pointedly identified the efforts of both for- and non-profit online peer markets and the challenges workers and market operators face [24, 8, 33]. Some researchers have implemented independent micro-work markets for their own purposes [7], though the purpose of this research was not to explore the idea of a worker-centric or worker-led market.

Computational social scientists have documented workers' efforts to circumvent the systems imposed on them by market operators [24], and more directly researchers have observed the continuing effort to resist and critique markets for "gig work," in these cases in the context of online labor, where micro-work on Amazon Mechanical Turk (AMT) predates offline gig work companies such as Uber [20, 30].

Nevertheless, frustrations with these marketplaces persist, and the trends among emerging marketplaces seem to commoditize workers more and more aggressively. These "patches" of existing markets appear to have only marginal effects on the qualities of these markets: Uber drivers continue to resist algo-rithmic matching while walking a fine line to avoid retribution from management, and some of the most frustrating requesters on AMT continue to antagonize Turkers.

**Juxtaposed, but not Synthesized**

Broadly, these bodies of research have overlapped in limited cases [30, 20]. The intersection of "community-driven action as a mode of designing" and "design of technologically enabled peer markets" represents a field site ready for the application of existing knowledge, and perhaps the production of new. We begin to bridge this research by applying our learnings regarding collective action (both offline and online) to what we have learned from research surrounding the "gig economy" in its various names. Drawing on the spirit of critical theory research and the grievances found among gig workers [20, 24], we began our fieldwork thinking about the viability of a self-directed crowd.

**METHOD & POSITION**

We approached this research with some amount of reflexivity, as described by Geertz and elaborated by others [14, 25], to acknowledge the role we play as participants in the culture we study. Reflexivity affords other benefits, however: it allowed us to avoid an undue and perhaps hopeless struggle to separate ourselves and objectify the communities we studied, and engage with participants more transparently, rather than attempting to abandon our preconceptions and existing mental models.

**Finding Participants, Making Partners**

Our process began by finding organizations with demonstrated success in areas such as worker advocacy and collective action, hoping that their experience in offline work would inform online working groups. We found a variety of organizations ranging small worker cooperatives to national-scale labor unions, and began to make contact with several. After a number of critical introductions by mutual contacts, we had a diverse range of partners ranging in size and spanning several industries.

We worked with five organizations: Organization A, organization B, organization C, organization D, and organization E. Organization A is a labor union that spans the nation, although we specifically worked with a union chapter local to our area; workers in this group are regulated by state laws describing qualifications and eligibility to work. This represented one of the more formalized, organized groups that we worked with. Organization B represents more than one million workers across various industries. The specific vertical of work we studied involved health care, and the degree of regulation involved with this group was even greater than with organization A. Organization C is a national labor union and worker advocacy group which supports tens of thousands of workers, many of whom are women and of minority groups. We worked with a local group that organization C supports named organization E, which helps domestic workers and house cleaners find work in our approximate geographic area. Finally, we worked briefly with organization D, a labor union that broadly advocates on behalf of many organizations of workers and expressed interest in representing existing workers in the "gig economy," such as Uber and Lyft drivers.

With each group, we explained our interest and motivation as researchers exploring the idea of a worker-centric labor market; we also attempted to describe what we hoped to contribute to this space. At the time, we imagined a "worker cooperative" —a worker-owned organization that often makes decisions democratically—and we found surprising (at the time) resistance from some worker-led organizations with which we spoke. But this presented us with opportunities to learn about the perceived shortcomings of various organization structures.

As our interviews and discussions progressed, we mocked up designs based on what we learned. By bringing these mockups to subsequent meetings, and iterating quickly based on the feedback workers provided us with, we were able to demonstrate in our own ways as computer scientists and designers that we were both listening to and learning from their input. This process, time-consuming and labor-intensive though it was, made important strides to prove the overarching claim that we were invested in these groups and respected the guidance they provided as equal partners.

Mock-ups offered another benefit that may be uniquely accessible to designers. As we found ourselves struggling to explain concepts such as privacy and trustworthiness in tangible ways, and thus failing to get substantive answers to questions around these issues, we realized that we could illustrate various paradigms, extrapolate from them, and learn how workers felt about design choices through their preference and especially from their suggestions for revision.

We ultimately spent more than three months making contact with, and learning from, these groups. For several weeks (admittedly fleeting, in the context of ethnographic fieldwork), we learned through participation–observation: answering phone dispatch requests and triaging the worker assignment system when interactions mediated by existing technologies broke down. We gained significantly deeper insights about workers and customers as a result of this time, but engaging in the uncomfortable, time-intensive, and tiring work of liaising with customers and triaging worker dispatch proved our level of commitment as greater in scope than a short-lived research project.

**EMERGENT THEMES**

We identified several aspects of the on-demand markets we studied that seem to apply in generalizable ways. These characteristics and issues extend beyond the more narrowly scoped issues of scheduling workers and handling payments which might seem more at the focal point of designing a labor market, although those issues too must be addressed; we focus specifically on the social negotiations that a technological system must broker. These guidelines are illustrated using mock-ups and a mobile application front-end developed in tandem with organization C, organization B, and organization E, and consist of the following considerations:

1. Constructive feedback
2. Assigning work fairly
3. Managing customer expectations
4. Protecting vulnerable workers
5. Reconciling worker identities

6. Assessing worker qualifications
7. Communicating worker quality

## 1. Constructive Feedback
*Ratings cause anxiety, whereas feedback can satisfy the same administrative needs without causing workers undue stress.*

At organization E, we learned that workers interpreted negative feedback very personally, making it difficult for them to internalize that feedback constructively and act on the suggestions customers made. Multiple issues may have been at play here: cultural differences might explain a mismatch in how feedback is offered and how it is received, but so too would the power imbalance between undocumented domestic workers and their customers.

Organization E didn't seem to investigate the cause of this effect, but their solution circumvented this problem entirely. Rather than giving workers unfiltered, raw feedback as it streamed in, organization E intercepted reviews from customers and distilled them into constructive, actionable feedback. Both praise and criticism would sometimes be read aloud to the entire group, with identifying information removed.

The purpose of reading positive feedback publicly was to reaffirm the group's sense of worth in a shared sense of success; the purpose of reading the negative feedback, meanwhile, was to prompt workers to reflect on what went "wrong," and how to avoid such an outcome in the future. A shared sense of failure also seemed to affect workers in these cases; whether this is an outcome of public readings of reviews, or more complex relationships between workers and organization E, is unclear.

Speaking with Uber and Lyft drivers, we found an alternative approach to providing workers with feedback: drivers are made aware of an aggregated rating—generally a moving average of the previous $n$ ratings—but they are not exposed to qualitative feedback in any form, even when customers provide it.

A driver's current aggregated rating, it turns out, is extremely important: a driver's rating determines the worker's eligibility to do work, and falling below various thresholds carries various consequences. Reports suggesting that these thresholds are finely tuned as well as closely guarded secrets make these markets particularly emotionally taxing for workers [11].

When we spoke to drivers, they relayed stories of apparent obligations to take expensive remedial courses if their overall rating dropped below a certain threshold, and widely held fears of suspension and deactivation for falling below an unknown level weigh heavily on workers' minds. Frustration directed at drunk, clumsy, or naive passengers accidentally or deliberately giving drivers a rating of 4 (out of 5) stars further exacerbates stress.

Markets which aim to empower "gig workers" should use quantitative rating systems sparingly, or to prompt more detailed qualitative feedback, as we illustrate in Figure 1. Quantitative ratings in the form of Likert scales followed by quali-tative prompts for feedback seem to precipitate more detailed feedback [17].

Figure 1. Given quantitative prompts and qualitative follow-up questions, research suggests that users are more likely to write more, which can better inform.

Quantitative metrics—especially those that are opaquely evaluated—do not benefit workers, nor does it seem they even inform appropriate customer behavior [32]. Meanwhile, constructive feedback suggesting improvements may provide workers with the necessary information to improve as professionals without causing undue stress over issues such as worker eligibility.

## 2. Assigning Work Fairly
*Who gets new work opportunities first is a contentious issue which should be negotiated by people, enforced by technology.*

When new work becomes available, who should have first access to claim it proves to be a contentious topic. Depending on the intended velocity of the market, various suggestions have emerged.

In existing technologically enabled markets, we find some of these decisions have been made carefully. Some of these approaches are worth considering, and can be described thusly:

- On AMT, Turkers find themselves in constant competition with one another to claim HITs (human intelligence tasks); the effects of this task searching behavior have been explored in some detail [10].
- Airbnb, Craigslist, and others invite workers to describe the services or products they offer, and leave the work of matching to customers. The complex nature of finding a suitable place to live may make this approach preferable over automated matching.
- Uber, Lyft, and others use algorithmic matching to determine who should be offered work; it is unclear to workers and consumers who is given a job offer first, but drivers we spoke to assumed that the nearest drivers were given first offer of a new job in markets such as Uber & Lyft.

Among each of these approaches designers can make choices that dramatically affect the experiences for workers and customers. The first option, used by AMT, Craigslist, etc. makes it possible for job turnover to be extremely high, but the cost is that workers may feel stressed by the uncertainty over whether a job they're considering has already been claimed. The second approach mentioned, which Airbnb uses, gives customers the choice to contact "workers," but the process of finding a provider makes the overall matching process time-consuming.

When we spoke with members of organization B and organization A, the consensus we initially found was that seniority within a reasonable proximity to the work would be the fairest solution to the problem we described. After some discussion, it emerged that this position was not universally shared among members of either group. Perhaps understandably, younger members of organization B felt that workers who had been working longer that day without having been offered any jobs, or workers who were significantly closer, should eventually be given higher priority for work.

Organization E chose yet another option, primarily using a random selection process at the beginning of each hiring day by pulling the names of currently present workers out of a jug: every eligible worker has a reasonable chance at being selected first, just as they have a chance of coming up last. Workers could short-circuit the selection process by volunteering to do work beneficial to the group—for example, making coffee, or sweeping the common room's floors. By volunteering to do this work, they're guaranteed first priority for incoming jobs the following day.

The best option for most markets might be the model Uber, Lyft, and others have adopted, where customers are automatically routed to workers, who are given a time window where they have exclusive claim to a job that's been posted. If they reject a job offer, the job moves to the next eligible worker.

The task of finding eligible, willing workers can be simplified by considering availability, narrowing the list of candidate workers to those who are interested in working during a given time frame. This can, however, be a particularly frustrating challenge for on-demand workers. One of the most common reasons we heard from gig workers who preferred such work was the flexibility that mode of work offered; in other words, trying to get workers to commit to windows of availability goes against the nature of that work—the freedom—which made it appealing to those we consulted.

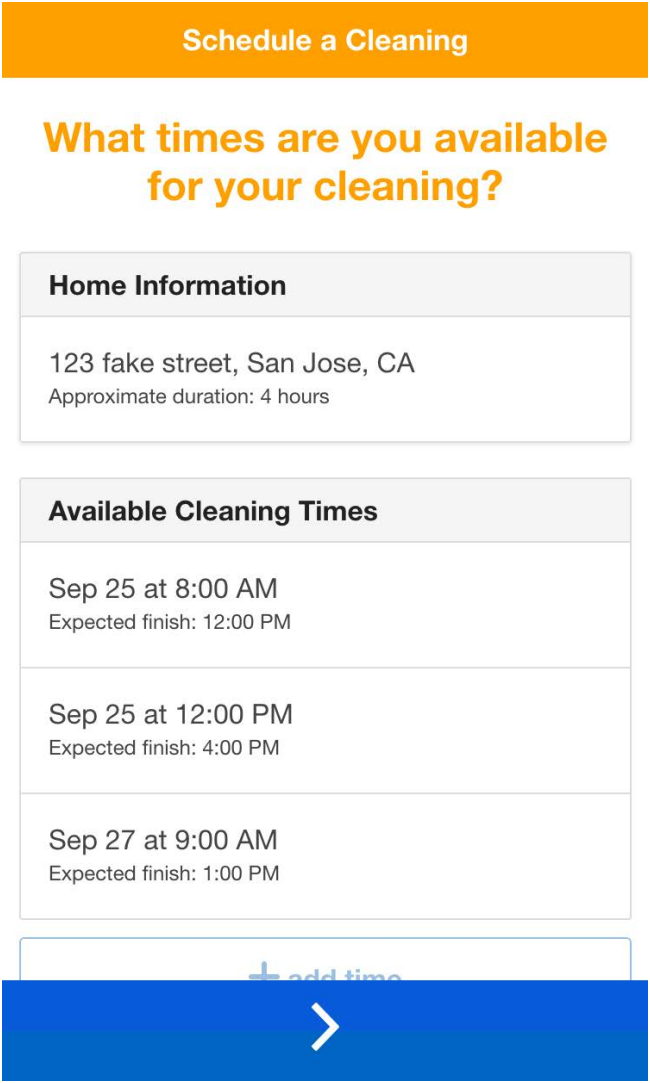

Figure 2. From the client's perspective, a lightweight scheduling system can suffice to communicate availability.

This challenge was ameliorated in various ways across field sites. At organization E, workers were assigned each day even if work was solicited days or weeks in advance (unless, of course, a specific worker was requested for repeat work). This process neatly handles the potentially ambiguous question of availability for a number of workers, but it's limiting in two ways:

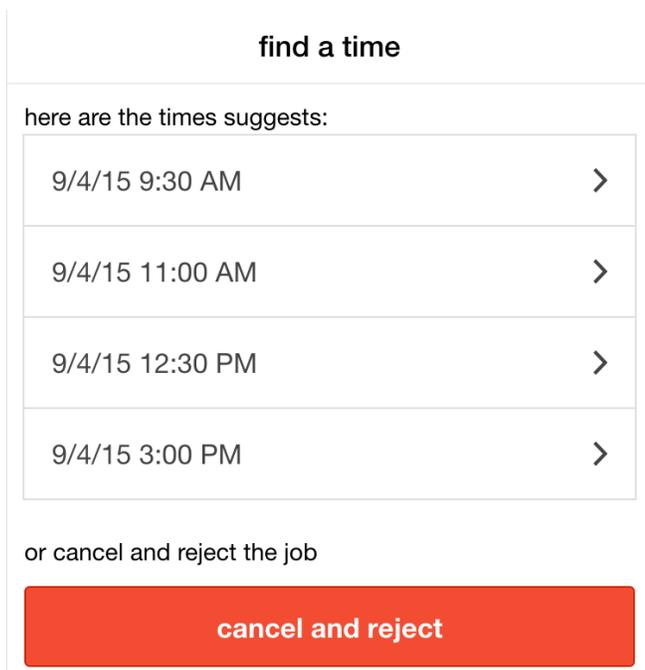

Figure 3. From the worker's perspective, scanning suggested start times can make scheduling easier than entering a series of windows of availability each week or month.

1. Organization E's scale is ultimately limited by their ability to process workers on a daily basis (this was particularly salient as a volunteer was processing workers as they arrived to check in at 7 A.M.).
2. Workers who wished to be available for leads had to travel to the organization E offices; sometimes workers would travel north for 30 minutes, only to get a job 30 minutes south again.

Instead of setting availability in advance, we suggest optimizing the process of scheduling so that workers and customers spend a minimal amount of time negotiating one another's availability.

This leaves open the question of how that list of workers is determined: who is the first choice, and who's second if the first worker doesn't claim the job? Who ends up getting last pick?

We can only offer that this factor is determined in large part by the organization running the system. A template for a worker-run marketplace should therefore expose these options, and more importantly surface the debate at the core of these discussions, while abstracting away the technical details such as implementation. These cases also illustrate that systems can be designed in ways that meaningfully and fairly reward workers for doing necessary work for the benefit of the community as a whole.

### 3. Managing Customer Expectations
*Clear expectations of all participants in trade benefit all, but introducing those expectations can be difficult for all parties.*

Perhaps the most common source of dissatisfaction with gig workers comes from a miscommunication or simply a mismatch in the expectations of the customer and the worker. The effects of this mismatch appear to have been considered when creating AMT, leading to norms where "good" requesters know to provide examples of correct work and highlight common mistakes, explicating the intent of the micro work. On other platforms, clear expectations are established ahead of time (on ride-sharing markets, the expected driving path, estimated time, and estimated cost is displayed to customers).

People we spoke with at organization B, organization E, and organization C felt that formalizing the expectations of workers was the most appropriate solution to this problem. By making it clear to customers what workers are expected to do (for instance, what specific tasks go into cleaning a kitchen), expectations of customers are set reasonably, workers can be trained properly, and disputes about whether the worker did the job properly or sufficiently become less ambiguous.

We tend to agree with this approach. In our fieldwork we found that paper checklists made it clearer to customers what to expect to be done—and importantly, what to expect not to be done. If customers wanted additional work completed, they could specify that and agree to the extra time it would take to do that job, which they sometimes did.

This "contract" between workers and customers can be written collaboratively by both parties, outlined by the workers themselves, or directed principally by customers as they describe the work they need completed; the exact process is less important than the shared understanding of the constituent tasks.

### 4. Protecting Vulnerable Workers
*Many workers in the gig economy are vulnerable to exploitation; careless or malicious systems can endanger workers.*

At organization E, workers expressed concern when we discussed the ability to show a profile picture to customers in advance of their arrival. These concerns stemmed from fears of their privacy being violated, and discrimination over their race, age, gender, and other characteristics. In particular, workers at organization E were afraid that customers would reject them for being too old.

Organization E addressed this issue by emphasizing the qualifications of workers, abstracting their names and details for first-time jobs. Workers fundamentally have final say over whether to be matched to a customer, based on general data about the location, an indication of whether they've worked together before, and a rough estimate of the amount of work solicited. (In most cases, a precise estimate of how long a job will take is difficult to make without someone familiar with the work on-site to make an informed estimate.)

If a customer has hired through organization E before and wants to hire a previously hired worker, a rudimentary system allows volunteer dispatch staff to identify those workers for matching. This allows workers to gradually develop a body of clients over time; eventually, they no longer rely on organization E to refer customers to them at all. Organization

Figure 4. Explicit checklists make it easier for workers to verify that they've completed all expected work; it also makes it apparent to customers if they need to add or clarify certain tasks.

A utilized a similar approach, allowing customers to request a worker they've worked with in the past, but otherwise not revealing information about workers to customers.

Many of the existing marketplaces for gig work don't expose such affordances to request a previously hired worker. This may be deliberate, as it effectively prevents workers from fostering a community of clients. By preventing workers and customers from becoming familiar with each other, workers remain dependent on the market created by the platform to provide leads on customers. Whether a system promotes such dependence or not is up to the system's designers, but we articulate a possible approach which makes repeat work possible.

As we illustrate in Figure 5, a simple interface exposing basic information about a previously hired worker can remind a customer whom they want to select. Further, a technological solution such as this can ease the negotiation of scheduling: in our case, we suggest prompting the worker for their availability; once several times are suggested, the customer can select and confirm a time.

Figure 5. By exposing some information about previously matched workers, a system can facilitate workers developing a reliable customer base.

Working with organization B, nurses we spoke to expressed concern over disrespectful or uncooperative patients. Nurses described various safety concerns including violent neighborhoods, dangerous pets such as dogs, and patients in declining mental health. Nurses told us that these were inherent risks associated with in-home care. Some of the nurses that we spoke to clarified that, in cases where they felt a particular danger, they may decline to meet a patient and communicate that concern later.

Technological systems can assist workers in cases like these: by exposing customer contact information to workers (and potentially vice versa), workers can contact customers before they begin work to clarify or resolve any issues.

### 5. Reconciling Worker Identities
*The very features that make gig work appealing also stymie attempts to coalesce stable worker advocacy groups.*

At organization E, organization A, organization C, and organization B we observed a subtle but important characteristic: everyone at these organizations identified themselves by the work that they did. Members of organization E self-identified as house cleaners and day laborers; people at organization A were workers, first and foremost. The same fundamental sentiment was shared by nurses affiliated with organization B.

We found that this carried an important effect. At organization E the effect it was most noticeable that workers identified together and recognized the importance of maintaining or-ganization E's positive reputation among customers. Every week, organization E administrative staff would read reviews of workers, made anonymous beforehand, allowing the workers to reflect on these frequent successes and occasional failures.

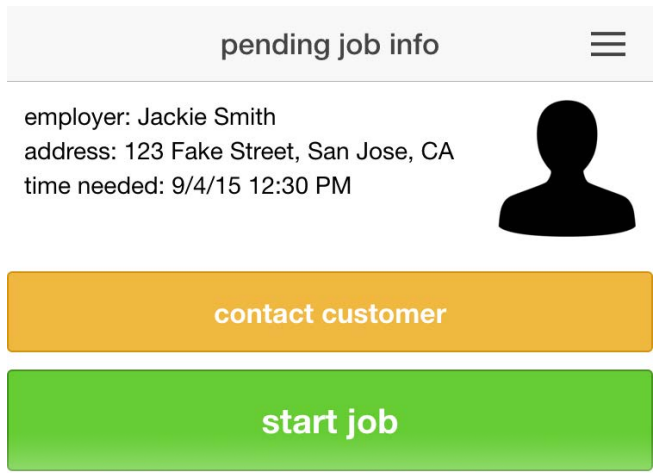

Figure 6. A simple interface for workers, allowing them to contact customers, facilitates the resolution of potential issues before work begins, such as difficulty finding a location, ambiguity in instructions, etc.

Perhaps most surprising was a shared sense of success and failure when workers listened to this feedback; people we spoke to felt that they had upheld or let down their peers and the organization of which they were part, even if the review was not of their own work. This strong sense of group identity seemed to drive workers to work hard to maintain the already good reputation that organization E enjoyed.

When we spoke to drivers in the gig economy, some told us that they had previously worked for competing yellow cab companies. They explained that the reason they gave up on conventional cab companies was because requirements to drive for yellow cabs felt onerous. Drivers worked for Uber and Lyft because those markets allowed them to drive as much or, importantly, as little as they wanted. At some yellow cab companies, a monthly fee was required for access to the cabs themselves.

This presents a potential challenge for those hoping to build communities fostering collective decision-making and action. As Kraut et al. point out [22], a crucial component in designing successful online communities is identification with the group. This is already a difficult goal to achieve, now exacerbated by the nature of the community involved. Features that make gig work appealing—the freedom to come and go as one pleases, for instance—make it that much more difficult to form a shared sense of community based on shared identities.

Indeed, workers in the gig economy tended not to think of themselves as identified by the work that they do— many drivers, for instance, volunteered that they were musicians, security guards, and even elementary school teachers. Their work as drivers, then, was not substantively important to who they are. Workers at organization E, meanwhile, identified as cleaners, day laborers, etc. without qualification.

Among many of the "gig workers" that we spoke to, the sense of freedom and independence turned out to be an important feature of their identity. The relative freedom over when and indeed whether to work was, it seemed, powerfully appealing.

This represents a difficult dilemma for labor unions and other conventional worker advocacy groups, which rely on the shared sense of identity of workers, who are now increasingly detaching their identities from the work that they do.

We assume that a worker-centric market needs emotional investment from its participants in order to succeed, and the evidence seems to corroborate this sense [30, 27, 16]. Given this premise, it seems that workers in a system must appreciate their reliance on their group's collective success. Balancing this need with the desire that workers have expressed to remain unburdened by various requirements might prove to be a difficult act, but a necessary one.

We have little concrete guidance for this challenge except to say that system designers might strive to find ways to make it palatable for gig workers to identify collectively and to recognize that their individual success is collectively determined by the quality of all of their work. By forming a sense of shared identity and investment in that community, a constructive sort of self-policing might emerge, as briefly discussed by Lave & Wenger [23].

## 6. Assessing Worker Qualifications

*Workers in variably regulated markets similarly seek to prove their qualification; they accomplish this in different ways.*

In interviewing workers across industries we found that workers with formalized qualifications wished to emphasize the value of the formalized qualifications that they offer; in the case of electrical workers, strictly regulated by the state in which we were conducting research, workers feared that unlicensed, unregulated workers potentially tarnished the reputations of workers with legitimate credentials.

In the driver-for-hire market, we generally trust that drivers for Uber, Lyft, and even yellow cab companies are legally qualified to drive a car. While passengers rarely ask to see proof of licence and it is rarely shown (except, sometimes, in the cases of yellow cabs), it is widely assumed that a worker given access to the market and its consumers has sufficiently proven to the market operators (Uber, Lyft, etc.) that the driver is qualified. The perceived risk of being caught without a valid driver's licence seems sufficiently discouraging that fears of abuse and deception are not widely held in the United States, where this research was conducted.

As we interviewed members of organization A, we discovered a complex web of laws describing the ratios of variously qualified workers and other workers on construction and other work sites. In short, a work site would be deemed in violation of construction regulations if too many inexperienced workers are working without sufficient more senior workers on the premises to oversee the work. This requirement proves chal-lenging to follow under the status quo as a worker calling in sick abruptly one morning may cause the work site to fall out of regulation.

Computer scientists may observe that this problem is simple enough to address, at least to augment a human-driven process of finding another worker with sufficient credentials to take the absent worker's place for the duration of the absence. We consider this one of the more formalized qualifications processes, and is roughly in line with the qualifications concerns we encountered when interviewing home health care workers through organization B.

We found less formally regulated, often homegrown, modes of qualifying workers at other organizations. At organization E and through organization C, we found that organizations would determine worker qualifications using highly specialized criteria. For instance, at organization E, we found that for day laborers they kept track not just of things like whether they were proficient English speakers, but also kept track of a wide array of qualifications describing whether they were eligible to take jobs involving yard work, heavy lifting, painting in various forms (e.g. with a brush versus with a spray canister), and numerous other factors.

This system generally worked, with a notable exception. One day, while volunteering for the phone dispatch system at organization E, we received a call asking for someone who could paint with a spray canister. We correctly matched a worker to that call, but when another worker arrived to pick the worker up (the cheapest of three options for getting the worker to the work site), it came to light that the worker did not want to work for a contractor—which this customer was. Our solution was to hastily find another worker who would be able to paint, which we managed to do expeditiously. Later, we discovered that we had neglected to verify that the worker was able to paint with a spray canister, which he was not. Ultimately, the worker was sent home and the work was not completed that day.

It should go without saying that this was a failure in attempting to match a worker and a customer. The individual points of failure, however, can inform the design of an automated system to match workers and customers. The first in the series of otherwise avoidable mistakes occurred as a result of the worker not being able to make an informed decision about the customer requesting them. When the problem emerged, we sourced a worker according to a different procedure from the method used during the rest of the day—specifically, hastily— and as a result we overlooked details that otherwise would have prevented the second worker from claiming (or even wanting to claim) the work for which he was not qualified. These considerations would be simple, but not necessarily intuitive, to implement in a computer system.

### 7. Communicating Worker Quality
*Groups use social pressure to encourage good actors or discourage bad actors & ensure high quality, but rarely both.*

Another issue that arose during fieldwork surrounded the verification that a worker was of high quality. To reference existing platforms again and using Uber and Lyft in this case, the quality of a worker can be illustrated simplistically by the quantitative ratings that drivers often have. The distinction between qualification and quality is important; to use driving as an example once again, a driver may be qualified (typically, licenced to drive & insured) but not high quality—for example, struggling to navigate in tricky roads, unfamiliar with the area, etc.

We found two primary ways of communicating worker quality:

- guaranteeing outcomes
- vouching for good work

At organizations with high barriers to membership, work can be guaranteed by the organization of workers. This guarantee gives customers relative confidence that any issues will be rectified at a cost fully absorbed by the organization, rather than the customer. To cover these costs, work groups generally tax workers for each job to form a "rainy day fund" in anticipation of such an expense.

In the case of organization A, if work is not done satisfactorily well, organization A's guarantee involves paying for a new worker to do the job correctly at the organization's expense. Workers whose work requires fixing are punished in various ways depending on the nature of the error and whether the worker has done inadequate work in the recent past; workers essentially face increasingly severe punishments as they repeatedly make mistakes (or make mistakes of greater cost).

It is important to note that these organizations can make this guarantee for two reasons. The first reason is that membership exposes workers to work opportunities that make attempting to gain membership worthwhile. The second reason is that membership eligibility is non-trivial to achieve, and transgressing community norms and risking expulsion carries significant consequences, especially given the work opportunities that they would be jeopardizing.

Organization E addressed this challenge using community pressure fostered by shared emotional investment in the organization. Through democratic administrative systems and frequent meetings mandatory for all active workers, a sense of communal buy-in seemed to take form among workers who felt that their failure to do good work would let down the rest of their peers. Organization E further stoked this sense of community investment by reading feedback from customers, made anonymous by organization E, to the group during all-hands meetings. Workers reportedly felt a sense of shared success when good feedback arrived, and similarly felt a shared failure when feedback was critical.

We discovered that quantitative ratings weigh heavily on workers, causing anxiety over arbitrary and opaque feedback negatively affecting their eligibility as workers. Platforms like Uber and Lyft warn their workers that a user rating below a certain threshold (for example, rolling average under 4.8 on a 1–5 scale) may result in disciplinary action. Drivers, for instance, told us of courses they had to take if the average of their ratings fell below 4.6/5.

We can—and will—discuss ways to make quantitative rating systems less damaging in practice, but we find that numeric ratings overwhelmingly tend to harm workers; feedback—rather than ratings—proves a more effective way to improve workers [17], and customers' responses to negative ratings seem to be exaggerated [32]. Given these apparent weaknesses in numeric ratings, we propose that efforts to communicate reputation should attempt to maximize qualitative data, rather than minimize it as many systems (e.g. Uber, Lyft, etc.) do.

This sense of shared reputation and the desire among individuals to uphold it was reportedly very strong. We discovered that workers, fearing their refusal to do certain kinds of work would reflect poorly on organization E, would agree to do work even when it was inappropriate to do so. Examples of impropriety include using unfamiliar mechanical equipment and working in hazardous settings with inadequate protection. In light of this, the organization has had to remind workers to take a firm stance when customers make unreasonable requests of them.

Whether worker motivation stemmed genuinely from investment in their community or from a more practical desire not to lose the job they've gotten that day is unclear. However, considering the high demand we found during our fieldwork we suspect that this practical desire was not significantly influential. In other words, workers could turn down work and confidently expect to be offered another worthwhile job, if they were only concerned with earning money.

## DISCUSSION

We spent several months working alongside workers in various industries, iterating on mockups and prototypes, and getting feedback from members and administrators of worker-advocacy groups like organization C, organization B, and organization D. With their insights, we identified a number of aspects of gig work that potentially marginalize workers, and in doing so we begin to articulate ways to avoid such an outcome. These potentially marginalizing aspects in the peer economy consist of the following:

1. Constructive feedback
2. Assigning work fairly
3. Managing customer expectations
4. Protecting vulnerable workers
5. Reconciling worker identities
6. Assessing worker qualifications
7. Communicating worker quality

We discussed the importance of such a system given the existing trends in technologically enabled labor markets, and suggest an alternative market design wherein workers might operate their marketplace collectively. Finally, informed by ethnographic fieldwork and interviews, we distill what we believe conscientious system-builders would need to create a worker-centric peer economy.

It is worth noting that while we worked with workers from various industries, seeking to find common threads between each community, we can't speak to the applicability of these findings across all conceivable industries and modes of work. Information workers such as those on AMT may differ in significant, fundamental ways compared to drivers-for-hire such as Uber drivers.

Above all, we hope to convey the potential insight one can gain by working with stakeholders, and the importance of collaboratively designing labor markets with the workers themselves.

As Kittur et al. point out, as researchers, computer scientists, and participants in the technological community we ask whether *"…we foresee a future crowd workplace in which we would want our children to participate"* [21]. Indeed, we must answer this question, by articulating an economy that empowers and respects workers, not one that than marginalizes and exploits them.